\documentclass{aa}

\def\drom{{\rm d}}

\def\rb{R_{\rm b}}
\def\pb{P_{\rm b}}
\def\msun{M_\odot}
\def\msuny{M_\odot~{\rm y}^{-1}}
\def\msuny9{10^{-9}~\msuny}
\def\mdens{\rm g~cm^{-3}}
\usepackage{graphicx}
\usepackage{natbib}
\usepackage{amssymb,amsmath}
\begin{document}
\title{Deformation and crustal rigidity
of rotating neutron stars}
\author{J. L. Zdunik\inst{1}  \and
M. Bejger\inst{2,1}
 \and
P. Haensel\inst{1}
}
\institute{N. Copernicus Astronomical Center, Polish
           Academy of Sciences, Bartycka 18, PL-00-716 Warszawa, Poland
{\em jlz@camk.edu.pl}} \institute{N. Copernicus Astronomical Center,
Polish Academy of Sciences, Bartycka 18, PL-00-716 Warszawa, Poland
\and LUTH, UMR 8102 du CNRS, Observatoire de Paris, F-92195 Meudon Cedex, France\\
{\tt jlz@camk.edu.pl, bejger@camk.edu.pl, haensel@camk.edu.pl
 }}
\offprints{J.L. Zdunik}
\date{Received xxx Accepted xxx}
\abstract{}{We calculate parameters $A$ and $B$ of the Baym-Pines model of the
hydro-elastic  equilibrium of rotating neutron stars. Parameter $A$
determines the energy increase of a non-rotating star due to a quadrupolar
deformation  of its shape. Parameter $B$ determines residual
quadrupolar  deformation due to the crustal shear strain
in a neutron star that spun down to a non-rotating state.}{The calculations of ${A}$
are based on precise numerical 2-D calculations for rotating neutron
stars with the realistic equations of state (EOSs) of dense matter.
An approximate, but quite precise, formula for $B$ is
used, which allows us to  separate the contribution of
the  crust from the dependence  on the
stellar mass $M$  and radius $R$. The elastic shear strain
distribution within the crust is modeled following
 Cutler et al. (2003). Realistic
EOSs of neutron star cores are used, some of them with
 high-density softening due to the
appearance of hyperons or a phase transition to an exotic state.}
{The values $A(M)$ and $B(M)$ were calculated for
$0.2\;M _\odot<M<0.9\;M_{\rm max}$
(where $M_{\rm max}$ is the maximum allowable mass) for seven
 EOSs of neutron star core,
 combined with several crust models. A standard formula based on the
 incompressible fluid model is shown to severely underestimate the value  of
 $A$. For $M<0.7\;\msun$ the values of $A(M)$ are nearly EOS-independent 
and are given (within a few percent) by a
 universal formula $A=3.87\;(M/\msun)^{7/3}\;\times
 10^{53}~{\rm erg}$.  We derive the
 scaling of $B$ with respect to $R$ and $M$,
 also valid for a thick crust. We show that $B$ for
 accreted crust strongly depends on pycnonuclear fusions
 at $\rho>10^{12}~\mdens$.
   }{}


\titlerunning{Deformation and crustal rigidity of rotating neutron stars }
\maketitle
%
\section{Introduction}
\label{sect:introd}
The solid crust of neutron stars forms quickly after their
birth. It plays an important r{\^o}le in neutron-star
dynamics. Apart from pressure, the stresses in the crust are
produced by an elastic shear strain; therefore, the crust can
have a non-axial deformation of mass distribution, which
allows a rotating neutron star to radiate gravitational waves. Moreover,
the inner neutron-star crust, with a crystal lattice of nuclei
immersed in neutron superfluid, can be a site of pinning the
neutron vortices to nuclei. A massive breaking of this pinning
after a maximal strain has been reached is thought to be
responsible for triggering the pulsar glitches.
Finally,
the presence of the crust makes a
free precession of a rotating neutron
star possible.

The hydrostatics of the neutron star crust is more complicated than
for an ideal fluid, because of the elastic strain that
produces an anisotropic contribution to the stress tensor.
To find the mechanical equilibria of a crust with
non-vanishing shear strain, one has to solve partial
differential equations of the hydro-elastic equilibrium, with
appropriate boundary conditions on the stellar surface and on
the crust - liquid core boundary.  A simple ``one-parameter
model''  of a neutron-star structure with crustal elastic
strain and its application to pulsar glitches was given by
\citet{BP1971}. The Baym-Pines model was reformulated in
the framework of General Relativity by \citet{CarterQuint1975}.
A neutron-star precession model was developed by \citet{PinesShah1972a,PinesShah1972b}, 
who used methods originally developed to
describe rotation of the Earth (\citealt{Munk.book} and
references therein). In these models, the neutron star is treated  as
a solid shell (crust) enveloping a liquid core.
More recently,  updated reviews of
models of precessing neutron stars was  presented by \citet{JonesAnderss2001}.
The crustal rigidity, resulting from elastic
strain in the deformed crust of neutron stars,  was recently
reexamined, using realistic dense matter models by
\citet{Cutler2003}.

The change in the stellar energy due to star's
quadrupolar deformation is
determined by the parameter $A$ of \citet{BP1971}. In the
present paper we calculate $A$ versus stellar mass $M$, using quadratic
expansions of {\it exact} quantities obtained in the precise
2-D simulations. To this aim, we have developed a systematic procedure
enabling us to calculate a quadratic response of neutron star
structure to a rigid rotation. Using a set of EOSs,
we study the EOSs dependence of  $A(M)$. Then we calculate the
parameter $B$ measuring  the elastic strain contribution to
the hydro-elastic equilibrium of a rotating star with crust.
In doing this, we avoid solving the complete hydro-elastic
equilibrium equations, by relying on results obtained by
\citet{Cutler2003}, and adapting them to different
EOSs of the crust and core. We then study
the scaling of $B$ with the mass
and the radius of neutron star and investigate
the differences between the ground-state and accreted crusts.

The plan of the paper is as follows. The model
of Baym and Pines is briefly described in Sect.\
\ref{sect:formulation}, where the parameters $A$ and $B$ are
defined. Calculation of $A$, based on the 2-D simulations for
rigidly rotating neutron stars with realistic EOSs, is
described in Sect.\ \ref{sect:alpha}.
The shear modulus of the  crusts
 is  briefly reviewed in Sect.\ \ref{sect:shear-crust},
 where we also  discuss the differences between the
 ground-state and accreted crusts. An approximation
suitable for calculating the crustal rigidity parameter, valid also
beyond the limit of a thin crust, is proposed in Sect.\
\ref{sect:crust.approx}. Existing calculations of $B$, and our
results for this quantity, are described in
Sects.\ \ref{sect:B}-\ref{sect:B-GS-ACC}. Scaling of $B$ with $M$ and $R$ is briefly
discussed in Sect.\ \ref{sect:B_MR}. Numerical results for crustal
rigidity versus stellar mass are presented in Sect.\
\ref{sect:num-res-b}. Discussion of our results and conclusions are
presented in Sect.\ \ref{sect:discussion}.
%
\section{The Baym-Pines model}
\label{sect:formulation}
We restrict ourselves to neutron star stars with an ellipsoidal
(quadrupole)  deformation of an axially symmetric  crust.
The rest mass of the star is
 fixed and  equal to $M_{\rm b}$. The stress tensor of dense matter
 is $\sigma_{ij}=\sigma^{\rm liq}_{ij}+\sigma^{\rm sh}_{ij}$,
 where the dominant, ideal liquid term
 $\sigma^{\rm liq}_{ij}=-P\delta_{ij}$, $P$
  is the pressure, and an elastic shear stress component
 $\sigma^{\rm sh}_{ij}$
 (see Sect.\ 5) exists only within the solid crust.
 Correspondingly, the energy of
 a configuration is $E=E^{\rm liq}+E^{\rm sh}$.

 Non-sphericity of an axially symmetric
 rotating configuration results in a moment of inertia tensor
 $\breve{I}_{ij}=diag(\breve{I}_1,\breve{I}_1,\breve{I}_3)$.
 The  quadrupolar deformation is fully described
 by the oblateness parameter
\begin{equation}
 {\epsilon}\equiv{{{I}_3-{I}_0}\over I_0}~,
\label{eq:eps.def}
\end{equation}
and diagonal elements of $I_{ij}$ are
\begin{equation}
{I}_1={I}_2=I_0(1-{\epsilon}/2)~, ~~~{I}_3=I_0(1+{\epsilon})~,
\label{eq:I_epsilon}
\end{equation}
so that the trace of ${I}_{ij}$ (invariant with respect to rotations
of the coordinate system) is $3I_0$.

 Our definition of oblateness parameter requires a comment. It
is the same as used by  \citet{Cutler2003}; however, it differs from
that of \cite{BP1971}, who defined it in terms of the crust  moment
of inertia, $\epsilon_{\rm BP}=(I_{{\rm c}3}- I_{{\rm c}0})/I_{{\rm
c}0}$. For stars made of incompressible fluid,
$\epsilon=\epsilon_{\rm BP}$; however, for realistic stars, made of
compressible matter, $\epsilon>\epsilon_{\rm BP}$.

 Two reference configurations are defined as follows.
 \vskip 2mm
 ${\cal C}_0$ - {\it non-rotating  and relaxed. } It is
 spherical, with $E^{\rm sh}({\cal C}_0)=0$ (no shear
 stress, completely relaxed) and has energy
 $E_0$. Its moment of inertia tensor is $I^{0}_{ij}=I_0\delta_{ij}$.
\vskip 2mm

$\breve{\cal C}$ - {\it rotating and relaxed.} It is rigidly
rotating, at  rotation frequency $\breve{\Omega}=2\pi/$(rotation
period). This configuration is also  completely relaxed,  so that
 $E^{\rm sh}(\breve{\cal C})=0$.

\vskip 2mm

An example of a reference configuration
$\breve{\cal C}$ is a newly born neutron star
that is already rigidly rotating but still completely liquid.
Then the star spins down, and its ellipticity
decreases to $\epsilon<\breve{\epsilon}$.
However, the crust solidifies quickly, and elastic strain
accumulates within the crust, which is opposing the decrease
in $\epsilon$.

Let us consider a rotating configuration ${\cal C}$,
with rotation frequency $\Omega$, and born as $\breve{\cal C}$.
Assume that there is a continuous spin-down
transformation $\breve{\cal C}\longrightarrow {\cal C}$ (no
crust breaking). Then, a ``residual oblateness'', which adds to that
resulting from rotation, is
\citep{BP1971,CarterQuint1975,Cutler2003}
\begin{equation}
 {\epsilon_{\rm r}}=b\breve{\epsilon}~.
\label{eq:eps.b}
\end{equation}
Here, $b$  is a ``rigidity parameter'' (called so by
 \citealt{JonesAnderss2001}), which depends on the elastic
properties of the crust, on the EOS of neutron star matter,
and on the mass $M$ of the reference configuration ${\cal
C}_0$. The name ``residual oblateness'' stems from
this being the ellipticity that remains  after  a complete spin down
to $\Omega=0$.

Within the Baym-Pines model, the value of $b$ can be calculated by
considering the limit of a non-rotating star. The quadratic
approximation to the stellar energy is then
\begin{equation}
E=E_0+ A\epsilon^2 +B(\epsilon-\breve{\epsilon})^2~,
\label{eq:E.res}
\end{equation}
where parameter $A$ is calculated from the quadratic
response of a non-rotating neutron star energy to a quadrupole
deformation. The last term on the righthand-side is $E^{\rm sh}$,
verifying $E^{\rm sh}(\breve{\cal C})=0$.
The equilibrium of a non-rotating star is reached at
minimum $E$, which implies
\begin{equation}
 {\epsilon}_{\rm r}={B\over A+B}\breve{\epsilon}~\Longrightarrow
 b={B\over A+B}~.
\label{eq:eps.A.B}
\end{equation}
In their calculation of $b$, \citet{Cutler2003} rely directly
on the  difference $\Delta I=I_3-I_0=I_0\epsilon$, calculated in two
cases. The value $\Delta I_\Omega=I_0{\breve{\epsilon}}$
is obtained for a fully relaxed configuration rotating at $\Omega$,
while the residual $\Delta I_{\rm d}=I_0\epsilon_{\rm r}$
results from a continuous spin-down from $\Omega$ to zero. Within
the quadratic approximation in $\Omega$, one gets $\Delta I_{\rm
d}/\Delta I_\Omega= b=\epsilon_{\rm r}/{\breve{\epsilon}}$.

Early calculations for a $1.4~M_\odot$ neutron star  have given
$A\sim 10^{53}~$erg and  $B\sim 10^{48}~$erg \citep{PandhapPS1976},
so that $b\simeq B/A\simeq 10^{-5}$.
 Recent calculations for a $1.4~M_\odot$ neutron star
with realistic EOSs of neutron star crust and core
 yields a much lower  value $b=(1.7-2.5)\times 10^{-7}$
 \citep{Cutler2003}. Dependence of $A$ and $B$ on the
 stellar mass was studied by \cite{Carlini1989}. As we will
 see, the approximations used by \cite{Carlini1989} were
 rather unrealistic.

\section{Rotational deformation of a relaxed configuration}
\label{sect:alpha}
We intend to calculate parameter $A$ as a function of the
gravitational mass of the non-rotating configuration of the
same baryon mass, $M_{\rm b}$. Consider configuration rotating
rigidly at angular frequency $\Omega$, as measured by a distant
observer. In what follows, we use the results of exact
2-D calculations. Einstein equations for stationary axi-symmetric
space-time  are solved using the {\tt rotstar} code, a part of
the {\tt LORENE} public-domain C++ scientific library based on
spectral methods of solution of partial differential equations
(see {\tt http://www.lorene.obspm.fr/}).  The stellar angular
momentum is $J$ and mass-energy is $Mc^2$. In what follows, in
the spirit of the Baym-Pines model, we restrict ourselves
to a slow rotation limit; i.e., we keep only lowest powers
of $\Omega$. To extract relevant coefficients, 
 we have chosen a non-dimensional expansion
parameter $\overline{\Omega}$ equal to the angular frequency 
$\Omega$ divided by the angular frequency for a star rotating at 
10ms period,
$\overline{\Omega}=\Omega/\Omega_{10{\rm ms}} 
= \Omega/(2\pi\cdot 100~{\rm s^{-1}})$.
\subsection{Coefficients $\alpha$ of the
star's response to rotation}
\label{sect:alphas}
\begin{table}
\caption{Equations of state of neutron star core.
}
\begin{center}
\begin{tabular}[t]{c|c|c}
\hline\hline EOS & model & reference\\ \hline BPAL12  & N
energy density functional  & a\\
FPS     & N
energy density functional  & b\\ GN3
& N relativistic mean field   &  c\\ DH    & N
energy density functional    &  d\\ APR   & N
variational theory     & e\\
BGN1H1 & NH, energy density functional    & f\\
GNHQm2 & NH + mixed baryon-quark state  &  g\\
\hline\hline
\end{tabular}
\end{center}
{Note: N - nucleons and leptons. NH -
hyperons, nucleons, and leptons.

References for the EOS:
 a - \cite{Bombaci1995}; b -  \cite{PandRavFPS1989};
 c - \cite{Glend1985}; d - \cite{DH2001};
 e -  A18$\delta$+UIX$^*$ model of  \cite{APR1998};
 f -  \cite{BG1997};  g - \cite{GlendBOOK}.
 }
\end{table}
\begin{figure}[h]
\centering
\resizebox{3.5in}{!}{\includegraphics[clip]{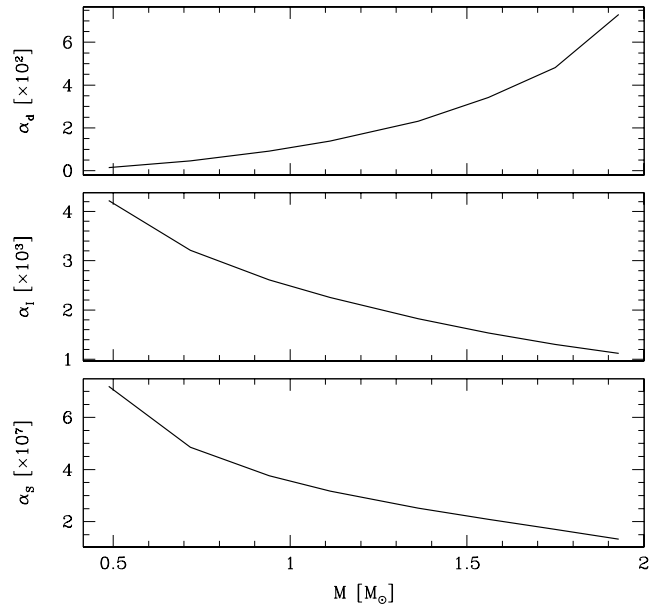}}
\caption{Parameters $\alpha_{\rm s}$, $\alpha_I$, and
$\alpha_{\rm d}$ versus stellar mass $M$, calculated for the
DH EOS (Table 1). For further explanations see the text.
}
\label{fig:param_SLy}
\end{figure}
\begin{figure}[h]
\centering
\resizebox{3.5in}{!}{\includegraphics[clip]{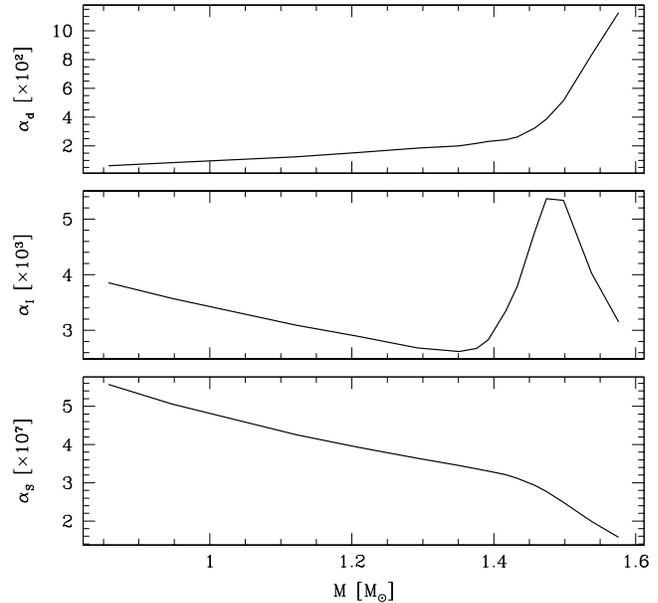}}
\caption{
Same as Fig.\ \ref{fig:param_SLy} but calculated for the
BGN1H1 EOS (Table 1).
}
\label{fig:param_BGN1H1}
\end{figure}
The non-sphericity of the mass-distribution  is measured by
ellipticity $\breve{\epsilon}$. We calculate and expand $I_3$
using
\begin{equation}
I_3={J\over \Omega}=I_0(1+\alpha_I\overline{\Omega}^2)~,
\label{eq:beta_I}
\end{equation}
which yields
\begin{equation}
\breve{\epsilon}=\alpha_I\overline{\Omega}^2~.
\label{eq:epsilon-beta}
\end{equation}
Simultaneously, we can parametrize the geometrical oblateness
of the shape of rotating star by
\begin{equation}
1-{r_{\rm pol}/ r_{\rm eq}}= \alpha_{\rm
s}\overline{\Omega}^2~,
 \label{eq:alpha_s}
\end{equation}
where $r_{\rm pol}$ ($r_{\rm eq}$) is the polar (equatorial) radial
coordinate of rotating star. Let us now consider the
decomposition of the total mass-energy
$Mc^2$ into the static part and kinetic energy.
The kinetic energy
is defined by analogy to the Newtonian theory \citep{FIP1986}
\begin{equation}
T={1\over 2}J\Omega = {1\over 2}I\Omega^2~.
\label{eq:T}
\end{equation}
Therefore,
\begin{equation}
Mc^2=M^{(0)}c^2+E_{\rm def}+T~,
\label{eq:E_def}
\end{equation}
where $M^{(0)}c^2$ is mass-energy of non-rotating
configuration, and $E_{\rm def}$ is the increase in
intrinsic energy of the star (i.e., calculated
in the star's reference system), as compared to the
non-rotating (static)   case.
To the lowest order, the
latter quantity can be parametrized as
\begin{equation}
E_{\rm def}=
\alpha_{\rm d}\left(1-{r_{\rm pol}/ r_{\rm eq}}\right)^2~.
\label{eq:alpha_d}
\end{equation}
The parameters $\alpha$ were calculated,
using exact 2-D models of rotating neutron stars,
for several realistic EOSs of neutron star cores. These EOSs
are listed  in Table 1. As the EOS of neutron star crust, we
used the model of \cite{DH2001} (hereafter referred
to as DH).
Examples of the dependence of the $\alpha$ parameters on
the gravitational mass of neutron star
are shown in  Figs.\ \ref{fig:param_SLy} (DH EOS)
and \ref{fig:param_BGN1H1} (BGN1H1 EOS).
The DH EOS assumes that the matter
is composed exclusively of nucleons and leptons, and
it leads to a monotonous behavior of all $\alpha$-s. Namely,
$\alpha_I$ and $\alpha_{\rm s}$ decrease with increasing $M$.
Such behavior is quite natural, because $\alpha_{\rm s}$ and
$\alpha_I$ measure the response of the shape and mass
distribution to a rotation at a fixed $\Omega$. This response
(polar flattening, anisotropy of the moment of inertia
tensor) weakens with increasing mass, because gravitational
pull, which tends to concentrate the matter towards the center,
 becomes stronger. On the contrary, the increase in
energy, relative to the spherical shape, due to a given
ellipticity of the star, is measured by $\alpha_{\rm d}$.
This energy  rapidly increases with star mass:
gravitational binding of the star rises with $M$,
 and the star becomes stiffer to polar flattening.

 There are some bounds on $M$ resulting from the high-precision 
constraint imposed on the 2-D simulations.
 Calculations of $\alpha$-s had to be stopped at
 $M \simeq 0.9~M_{\rm max}$, because for $M\approx M_{\rm
 max}$, the closeness of relativistic instability makes precise
 2-D calculation of linear response to $\overline{\Omega}^2$
 very difficult. On the other hand, $M$ could not be too low,
 because for low $M$ the rotating star becomes very ``soft''
 to rotational deformation, which is reflected in very high
 values of $\alpha_{\rm s}$ and $\alpha_{_I}$. All in all,
 very precise calculation of $\alpha$-s could be done only  for
 $0.2~M_\odot\la M \la 0.9~M_{\rm max}$.
 \footnote[1]{Let us remind that non-rotating cold neutron
 stars $M_{\rm min}\approx 0.1\;\msun$ (\citealt{HZD2002})}

\begin{figure}[h]
\centering
\resizebox{3.5in}{!}{\includegraphics[clip]{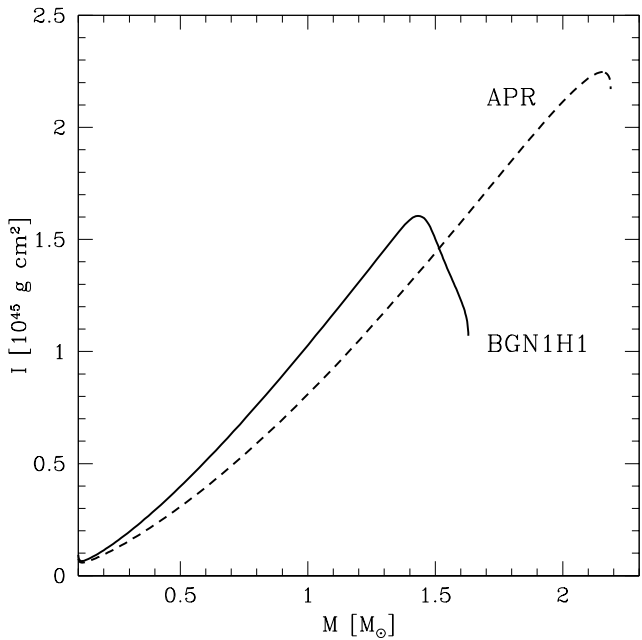}}
\caption{Moment of inertia for slow, rigid rotation, $I=I_0$,
versus stellar mass, $M$, for the APR and BGN1H1
EOS1 (Table 1).}
\label{fig:I_M_EOS}
\end{figure}

The BGN1H1 EOS takes the high-density
softening due to the appearance of hyperons into account. However,
for $M\la 1.35~M_\odot$ stellar core does not contain
 hyperons, and the $\alpha(M)$ dependence is monotonous,
 similar to the DH EOS.
 For $M>1.35~M_\odot$ hyperons soften
 the neutron star core, resulting in a specific
 $M(R)$ dependence. Namely, $M$ increases very weakly with
 a decreasing $R$ (flat $M(R)$ curve).
 In contrast, one notes a strong increase
  in the binding energy of the star, and a decrease in the
  moment of inertia with increasing $M$, Fig. \ref{fig:I_M_EOS}.
   The effects of the
  hyperon softening on $\alpha(M)$-s are
  therefore the following.
  The function $\alpha_{\rm s}$ ($\alpha_{\rm d}$)  decreases
  (increases) more steeply with
  $M$, because the star becomes more compact and its binding increases
  rapidly with $M$. The most characteristic, however, is the
  effect on $\alpha_I(M)$, which increases rapidly by a factor
  of two, reaching a maximum  at $M\simeq 1.5~M_\odot$, and
  then falls monotonously.
  This specific behavior is to some extent due to a
  rapid drop in $I$ for $M\ga 1.4~M_\odot$.

\subsection{Energy and oblateness - coefficient $A$}
\label{sect:A}
Using Eqs.\ (\ref{eq:epsilon-beta}), (\ref{eq:alpha_s}),
and (\ref{eq:alpha_d})  we obtain final
expression for $A$ in terms of three $\alpha$-s:
\begin{equation}
A= \alpha_{\rm d}\left(\alpha_{\rm s}/\alpha_I\right)^2~.
\label{eq:A-alpha's}
\end{equation}
%

\begin{figure}[h]
\centering
\resizebox{3.5in}{!}{\includegraphics[clip]{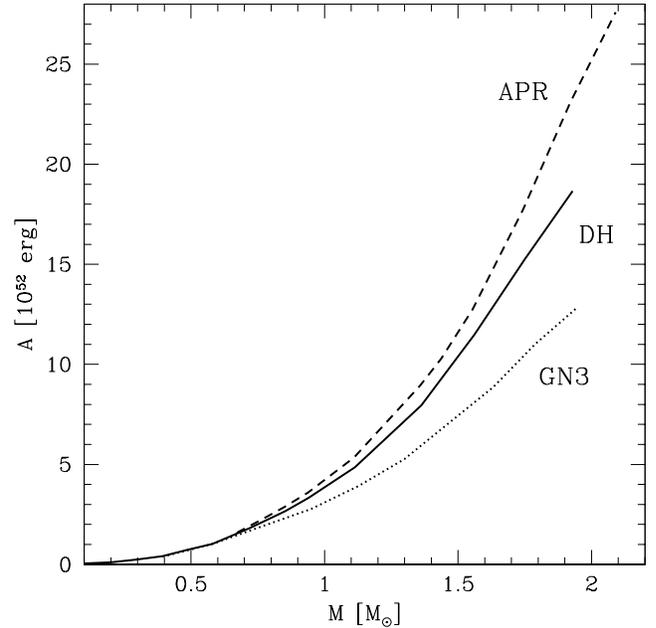}}
\caption{
Parameter $A$ versus stellar mass $M$ for the APR, DH, and
GN3 EOSs (Table 1). For more explanations see the text.
}
\label{fig:A_EOS1}
\end{figure}

\begin{figure}[h]
\centering
\resizebox{3.5in}{!}{\includegraphics[clip]{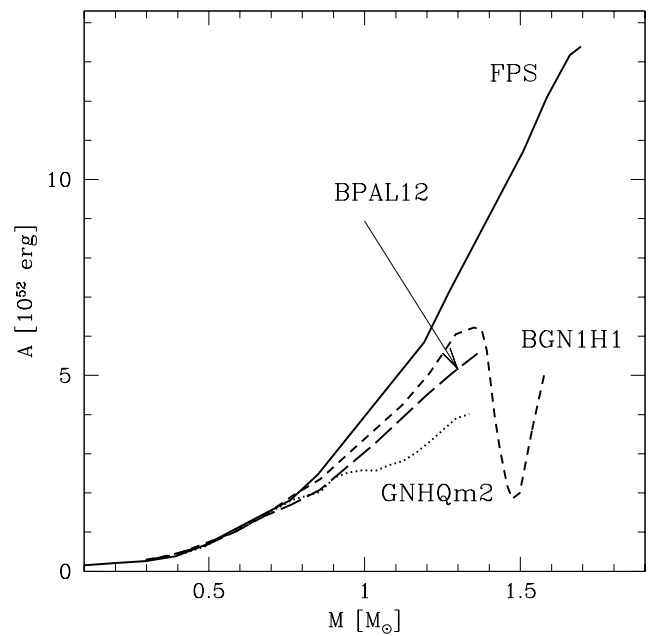}}
\caption{Same as Fig.\ \ref{fig:A_EOS1} but for
the BPAL12, FPS, BGN1H1, and GNHQm2 EOSs (Table 1).
For more explanations see the text.}
\label{fig:A_EOS2}
\end{figure}
\noindent Plots of $A(M)$ for seven realistic EOSs from Table 1
 are shown in Figs.
\ref{fig:A_EOS1} and \ref{fig:A_EOS2}.
As already explained in Sect.\ \ref{sect:alphas},
high precision 2-D calculations were only possible within a
limited range of stellar mass, $0.2~M_\odot\la M
\la 0.9~M_{\rm max}$. Notice that
EOSs in Fig.\ \ref{fig:A_EOS1} have no high-density softening.
They lead to monotonously increasing $A(M)$, with a maximum
scatter at given $M$ characterized by a factor of two.

In Fig.\ \ref{fig:A_EOS2} we compared $A(M)$ for two EOSs
with high-density softening, due to hyperonization (BGN1H1) and
to a transition to a mixed hadron-quark phase (GNHQm2), with
two EOSs without such a softening. Hyperon softening implies a
dramatic drop (by a factor of five!) for $M\ga 1.4~M_\odot$.
The appearance of a mixed-phase core leads to quasi-constancy of $A$
for BNHQm2 EOS near $M\simeq 1~M_\odot$.

\begin{figure}[h]
\centering
\resizebox{3.5in}{!}{\includegraphics[clip]{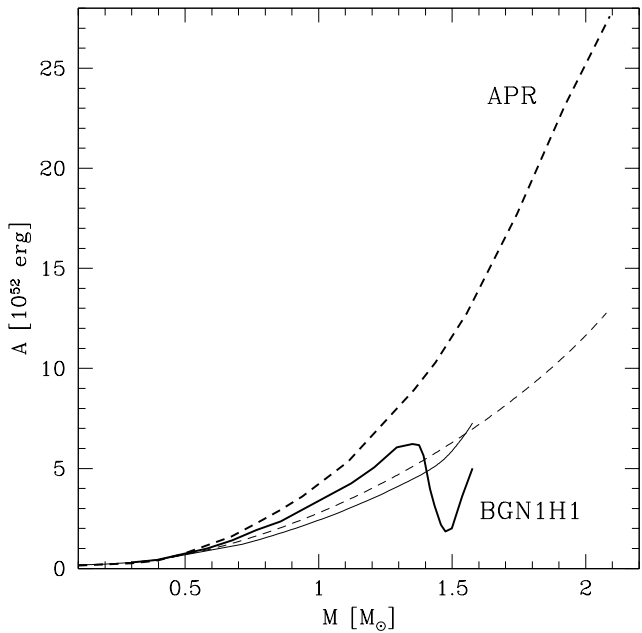}}
\caption{The coefficient $A$ versus stellar mass, $M$,
for the APR (thick dash line) and BGN1H1 (thick solid line)
EOSs (Table 1). Thin lines (dash - APR, solid - BGN1H1)
correspond to a standard incompressible-fluid
approximation, Eq.\ (\ref{eq:A_Love}). }
\label{fig:A_Love}
\end{figure}

Up to now, $A(M)$ has been calculated using the model of
incompressible-fluid stars in the Newtonian theory of
gravitation
(\citealt{BP1971,PandhapPS1976,Carlini1989,
JonesAnderss2001}). For such a model, the formula for
$A(M)$ was originally derived  by \cite{Love1920},
\begin{equation}
A={3\over 25} {GM^2\over R}~.
\label{eq:A_Love}
\end{equation}
As we see in Fig.\ \ref{fig:A_Love}, this formula badly
underestimates $A(M)$ for realistic neutron stars with a high-density
softening, and it is unable to
reproduce a strong drop, due to a high-density softening
of the EOS.

For a sufficiently low stellar mass, one can describe
the $A(M)$ relations by a universal formula,
because  the influence of the dense-matter EOS becomes
less and less important with decreasing $M$. Consequently,
the $A(M)$ relations for different EOSs converge to a
single one.
An approximate formula (fitted in the $0.1-0.7M_{\odot}$ range) reads
\begin{equation}
A(M) = a_1\;(M/\msun)^{b_1}\; 10^{53}~{\rm erg}~,
\label{eq:AMsmall}
\end{equation}
with $a_1=3.87$ and $b_1=7/3$. The fitting formula works very
well, with accuracy of the order of one per cent, for masses
less than $0.7M_\odot$. Its quality worsens for higher masses
when the $A(M)$ relations become EOS-dependent. Similarly, one
can approximate the numbers given by the
Newtonian Love's expression, where the true
values of $M$ and $R$ are used, by a universal function
 $A_{\rm Love}(M)$ analogous to Eq.~(\ref{eq:AMsmall}), but
 with $a_1=2.68$ and $b_1=2$. Such a formula for $A_{\rm Love}(M)$
 is quite precise (within a few percent) for masses lower
than $0.7M_\odot$.

\section{Shear modulus of the crust}
\label{sect:shear-crust}
We follow the notations of \cite{NSB1}.
Deformation of the crust with respect to the equilibrium
(ground state) configuration of ions (nuclei) in the lattice
implies displacement of ions into new positions
$\pmb{r}^\prime = \pmb{r}+\pmb{u}$, where
$\pmb{u}=\pmb{u}(\pmb{r})$ is the displacement vector (field).
In the macroscopic (continuum) limit, $\pmb{r}$ and $\pmb{u}$
are treated as continuous fields. Nonvanishing $\pmb{u}$
induces deformation energy density ${\cal E}_{\rm def}={\cal
E}(\pmb{u})-{\cal E}_0$, where ${\cal E}_0$ corresponds to the
non-deformed (equilibrium) crust. Non-zero $\pmb{u}$ induces
forces that  tend to return a matter element to the state
of minimum energy density ${\cal E}_0$. These forces result
from the elastic strain. The deformation energy is determined
by the strain tensor,
\begin{equation}
u_{ik}=u_{ki}={1\over 2}\left(
{\partial u_i\over \partial x_k} +
{\partial u_k\over \partial x_i}
\right)~,
\label{eq:u}
\end{equation}
and the elastic coefficients of the crust.  For an ideal bcc
lattice, there are two  independent elastic shear  moduli
$b_{11}={1\over 2}(c_{11}-c_{12})$ and $c_{44}$ (see, e.g.,
\citealt{Kittel1986}). For a classical Coulomb lattice, they were
first calculated by \cite{Fuchs1936} (see also
\citealt{Ogata1990}): $b_{11}=0.0245 n_{_{\cal
N}}(Ze)^2/r_{\rm c}$, $c_{44}=0.1827 n_{_{\cal
N}}(Ze)^2/r_{\rm c}$. Crustal matter is likely to be an
isotropic bcc polycrystal, with a single effective shear
modulus $\mu$. The shear strain contribution to the stress
tensor is then
\begin{equation}
\sigma_{ij}^{\rm sh}=2\mu\;
\left(u_{ij}-{1\over 3}\delta_{ij}\sum_{k}u_{kk}\right)~.
\label{eq:sig.mu}
\end{equation}

A standard choice in the preceding
studies of elastic effects in the neutron 
star structure and dynamics was $\mu=c_{44}$
(\citealt{BP1971,PandhapPS1976,McDermott1988},
and references therein). This choice significantly
overestimates $\mu$. A correct calculation, involving
averaging  over rotations of Cartesian axes, was done by
\cite{Ogata1990}, and gives
\begin{equation}
\mu={1\over 5}\left(2b_{11}+3c_{44}\right)=
0.1199 {n_{_{\cal N}}(Ze)^2\over r_{\rm c}}~.
\label{eq:mu_eff}
\end{equation}
We rewrite this formula in a form suitable for calculations:
\begin{equation}
\mu=7.816\times 10^{28}\;
\left( {\rho_{13}\over A_{\rm c,3}}\right)^{4/3}
\left( {Z \over 40}\right)^{2}~{\rm erg~cm^{-3}}~,
\label{eq:mu_eff_num}
\end{equation}
where  $\rho_{\rm 13}\equiv\rho/10^{13}~{\rm g~cm^{-3}}$,
 $A_{\rm c}$ is the total number of nucleons in the
ion cell, $A_{\rm c,3}\equiv A_{\rm c}/10^3$,
 and $Z$ is the number of protons in an nucleus.
This is the formula used in the present paper.
It should be stressed that this formula, strictly speaking,
holds for the outer crust, where the approximation
of point-like nuclei is valid. However, for simplicity
we also use Eq.\ (\ref{eq:mu_eff_num}) in the inner crust,
neglecting the effect of finite nuclear
size and possible existence
of non-spherical  ``pasta nuclei'' (see the end fragment of
Sect.\ \ref{sect:GS-mu}).
\subsection{Ground-state crust}
\label{sect:GS-mu}
This is a standard approximation assumed in the calculations
of the neutron-star structure (see \citealt{NSB1} and references
therein). At each pressure, matter is assumed to realize the
minimum of the enthalpy per nucleon. This is equivalent to the
minimum of the energy per nucleon  at every density; however,
in contrast to the pressure that is continuous and strictly
monotonous in the stellar interior,  the density can suffer jumps.

The ground-state approximation may be good for isolated neutron
stars formed in core-collapse supernovae. Initial temperature
is there $\ga 10^{10}~$K and the matter is in nuclear  equilibrium. In a
few months, following neutron star birth, the temperature falls to  $10^9~$K, so that
layers of $\rho\ga 10^8~\mdens$ are strongly degenerate. It is
tacitly assumed that  the matter  cools, keeping
nuclear equilibrium. As
a result, the cooled crust reaches its ground state.  Its
structure and the EOS can be calculated assuming $T=0$.

 In the inner crust, at
densities approaching $10^{14}~{\rm g~cm^{-3}}$, the nuclear
radius becomes comparable to the distance
between nuclei. The  effective nuclear charge becomes smaller
than $Z$ due to electrons present in nuclear volume, and
moreover the nuclei can lose their spherical shape, being
replaced by ``nuclear pasta'' (for a review, see \citealt{NSB1}).
However,
the very existence
of a ``pasta nuclei'' layer near the
bottom of the crust is model-dependent (\citealt{NSB1}).
The pasta layers  correspond to various phases
of liquid crystals and Eq.\ (\ref{eq:mu_eff_num}) is
not valid for them. Elastic moduli of some pasta phases
were calculated
by \cite{PP1998} (for a review, see \citealt{NSB1}).

\begin{figure}[h]
\centering
\resizebox{3.5in}{!}{\includegraphics[clip]{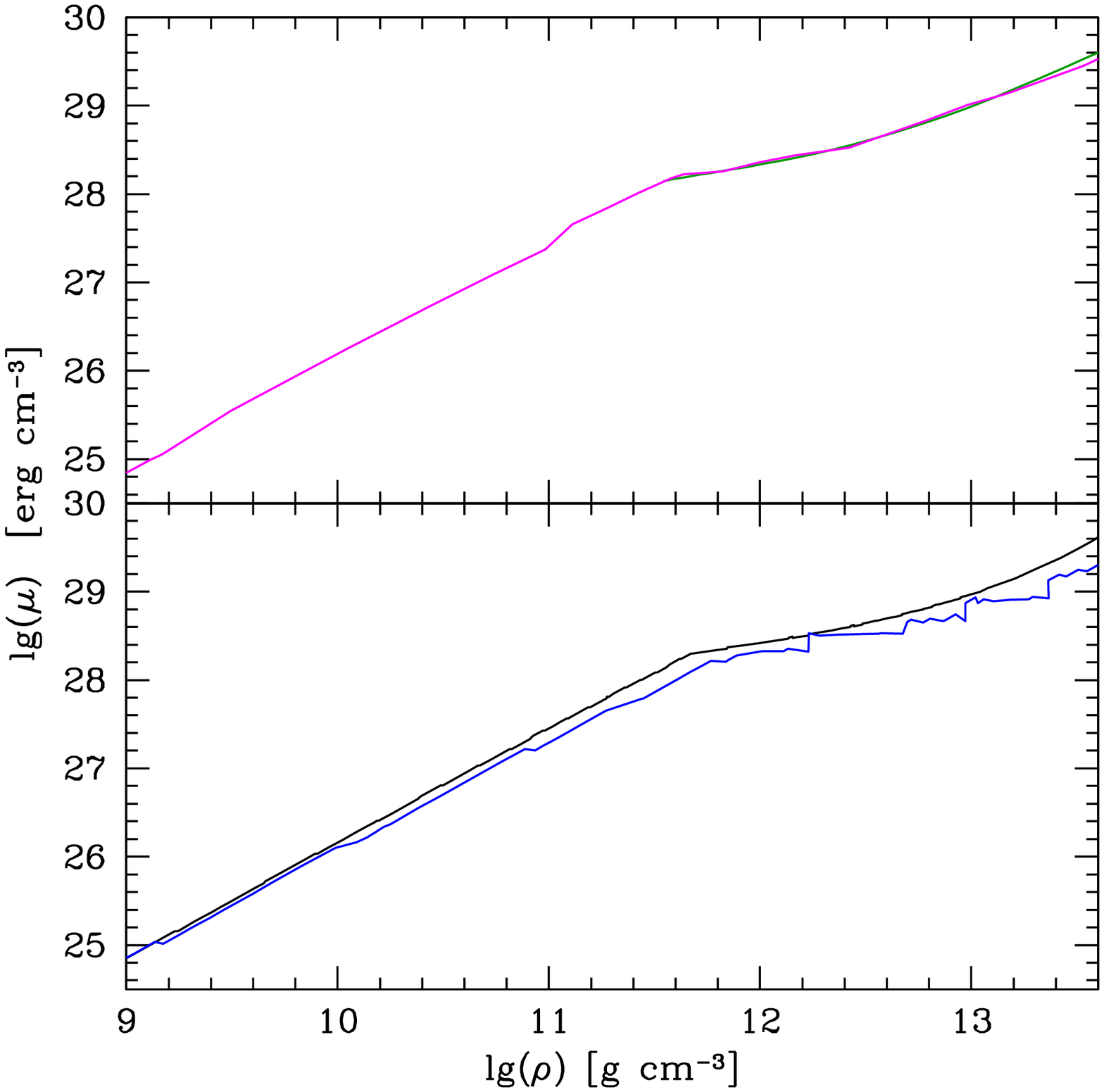}}
\caption{(Color online) Effective shear modulus $\mu$,
Eq.\ (\ref{eq:mu_eff}),
versus  $\rho$, for different models of the crust.
{\bf Upper panel:} ground-state crust. Magenta line - \cite{DH2001};
green line - \cite{NV1973}, only inner crust. Discontinuities
in $\rho$ have been smoothed via interpolation.
{\bf Lower panel:} blue line - an accreted crust model
 including pycnonuclear fusions (\citealt{HZ2008}). The ashes
 of X-ray bursts consist of pure $^{56}{\rm Fe}$, the crust structure
 given in Table A.3 of Appendix A of (\citealt{HZ2008}). Black
 line - ground-state crust, calculated using the same
 compressible liquid drop model of nuclei as the accreted
 crust.}
\label{fig:mu-GS-ACC}
\end{figure}

In the present paper, we limit ourselves to
 the crust models that do not exhibit pasta phases.
 We mainly use
the crust model of \cite{DH2001}, in which spherical nuclei
exist down to the bottom edge of the crust, so that Eq.\
(\ref{eq:mu_eff_num}) can be used as a reasonable
approximation. In the
upper
 panel of Fig.\ \ref{fig:mu-GS-ACC},
we show plots of effective shear modulus $\mu$ versus density $\rho$
for two models of the ground-state crust (\citealt{DH2001},
\citealt{NV1973}, the latter one for the inner crust only).
Differences between two models are small.
\subsection{Accreted crust}
\label{sect:accreted-mu}
Neutron stars that have gone through a long stage of accretion in
the low-mass X-ray binaries  have very different crusts from
the ground-state one. Accretion could last for as long as
$\sim 10^9$ years, with typical accretion rates $10^{-10} -
10^{-9}~\msun~{\rm y}^{-1} $.
After $\sim 10^7/(\dot{M}/10^{-9}~\msun~{\rm y}^{-1})$~years,
the original crust, formed at neutron star birth, is replaced
by an accreted one. During formation of an accreted crust,
temperature in the stellar interior with $\rho\ga 10^8~\mdens$
is at most a few times $10^8$~K (\citealt{Miralda1990}), and the
only reactions in compressed matter are those not associated
with overcoming the nuclear Coulomb barriers. Before the
neutron drip point at $\rho\approx 6\times 10^{11}~\mdens$,
evolution is mediated by the electron captures. At higher
density, neutron emission and absorption also play
important r{\^o}le. Finally, at $\rho \sim
10^{12}-10^{13}~\mdens$, the charge of proton clusters
becomes so low that  it has been speculated that
their  zero-point vibrations  around the crustal
lattice sites could possibly allow for the pycnonuclear fusion
\citep{Sato1979}.

Accreted crust, produced in the scenario described above, has
composition dramatically different from the ground-state one.
In particular, nuclei in an accreted crust have much smaller $A$
and $Z$ \citep{Sato1979,HZ1990a,HZ2003}. Therefore, an
accreted crust is softer to an elastic deformation than is the
ground-state one. To correctly calculate the
difference between  shear moduli of  the ground state and of
an accreted crust, one has to use the same nuclear model. We
assume the  Mackie-Baym model of nuclei in dense matter
\citep{MB1977}, in the version used in
\cite{HZ1990a,HZ1990b,HZ2003}.

Our results are presented in the
lower panel of
Fig.\ \ref{fig:mu-GS-ACC}. At a given $\rho$,
nuclei in an accreted crust have
lower values of $Z$ and $A$ than those in the ground-state
crust. The values of $\mu(\rho)$ are correspondingly lower.
One notices sharp drops and jumps of $\mu$ of accreted crust.
The drops are due to electron captures leading to
$Z\longrightarrow Z-2$.
In the inner crust, electron capture
triggers emission of neutrons, so that both $Z$ and $A$
drop at the threshold  density.
At some densities above
$10^{12}~\mdens$, $Z$ may be low enough to allow for
pycnonuclear fusion, resulting in a significant jump in $Z$
and $A$, corresponding to a very large jump in $\mu$. However,
one has to stress uncertainties in the description of zero point
motion of proton clusters carrying with them bound neutrons,
in the dense superfluid medium of unbound neutrons. Therefore,
this part of the scenario of the formation of accreted crust is plagued by
huge uncertainties, and even doubts, about whether the pycnonuclear fusion
occurs at all.
 In contrast to the accreted crust, we get a smooth $\mu(\rho)$ for
the ground-state crust, calculated using the same compressible
liquid drop model of nuclei. Notice that our model has no
shell correction term.
\section{Crust structure: an approximation}
\label{sect:crust.approx}
In the simplest approximation, we consider a spherical neutron star
built of an ideal fluid. The hydrostatic equilibrium
 is then determined by the Tolman-Oppenheimer-Volkoff (TOV) equations
 (see, e.g., \citealt{ST1983})
\begin{equation}
{{\rm d}P\over {\rm d}r}=-{G(\rho+P/c^2)
(m+4\pi r^3P/c^2)\over r^2\,(1-{2Gm/rc^2})}~,
\label{eq:TOV1}
\end{equation}
\begin{equation}
{{\rm d}m\over {\rm d}r}=4\pi r^2\rho~,
\label{eq:TOV2}
\end{equation}
where $r$ is the circumferential radius and $m(r)$
the (gravitational) mass contained within the sphere of radius $r$.
The boundary conditions are $m(0)=0$ and $P(R)=0$, where
$R$ is the star radius. The star mass is $M=m(R)$.
The bottom of the crust is at
$r=R_{\rm b}$, so that the mass of the crust is $M_{\rm cr}=
M-m(R_{\rm b})$. The pressure at the bottom of
the crust is $P_{\rm b}=P(R_{\rm b})$.

Within the crust, $0<P<P_{\rm b}$, $R>r>R_{\rm b}$
and $M-M_{\rm cr}<m<M$. We limit ourselves to neutron stars
with $M_{\rm cr}\ll M$, so that $m\approx M$
and $4\pi r^3P/mc^2\ll 1$ within
 the crust. Moreover,
within the crust $P/c^2\ll \rho$. Under these conditions,
Eq.\ (\ref{eq:TOV1}) can be rewritten as
\begin{equation}
{{\rm d}P\over {\rm d}r}=-{G M \rho
\over r^2\,(1-{2GM/rc^2})}
\end{equation}
or, equivalently, as
\begin{equation}
{{\rm d}P\over \rho}=
-GM {\drom r \over r^2 (1-{2GM/rc^2})}~.
\label{tovcr2}
\end{equation}
Within the crust, i.e.,
for $0<P<P_{\rm b}$, we define a
dimensionless function of pressure
\begin{equation}
\chi(P)=\int_0^{P} {\drom P^\prime\over
\rho(P^\prime)c^2}~.
\label{chio}
\end{equation}
Notice that $\chi(P)$ is determined
solely by the EOS of the crust.
Using Eq.\ (\ref{tovcr2}), one
obtains
$\chi$ versus $r$,
\begin{equation}
\chi={1\over 2}\ln
\left[{{1-r_{\rm g}/R}\over
1-r_{\rm g}/r}\right]~,
\label{rap2}
\end{equation}
where $r_{\rm g}\equiv 2GM/c^2$. Defining $a=1-r_{\rm
g}/R$, we obtain $r$ within the crust as a function of
$\chi$,
\begin{equation}
r=r_{\rm g}/\left(1-a {\rm e}^{-2\chi}\right)~.
\label{eq:r-chi}
\end{equation}
We can now derive a useful formula for an integral
of any function $f$ over the crust volume. A proper volume
of a spherical shell between $r$ and $r+{\rm d}r$ is
\begin{equation}
\drom V = 4\pi r^2 {\drom r\over (1-r_{\rm g}/r)^{1/2}}~.
\end{equation}
Using Eq.\ (\ref{eq:r-chi}) we obtain
\begin{equation}
\int_{\rb}^R f \drom V = 8\pi R^3
(1-a)^3 a^{1/2} \int_{\eta_{\rm b}}^{1}
f  {\drom\eta\over (1-a\eta^2)^4}~,
\label{intc}
\end{equation}
where $\eta={\rm e}^{-\chi}$ and $\eta_{\rm b}=\eta(P_{\rm
b})$. Therefore, for a given function  $f(P)$ and for
a given value of $\pb$, it is sufficient to calculate
a function
\begin{equation}
{\cal Z}_f(a)=(1-a)^3 a^{1/2}\int_{\eta_{\rm b}}^{1}
f {\drom\eta\over (1-a\eta^2)^4}~,
\label{eq:I_f}
\end{equation}
and the value of the
integral of $f$ over the crust of any star
having the radius and the mass of the star $R$
and $M$ can then  be obtained using  Eq.\ (\ref{intc}).
In our case $f=\mu$.

\begin{figure}[h]
\centering
\resizebox{3.5in}{!}{\includegraphics[clip]{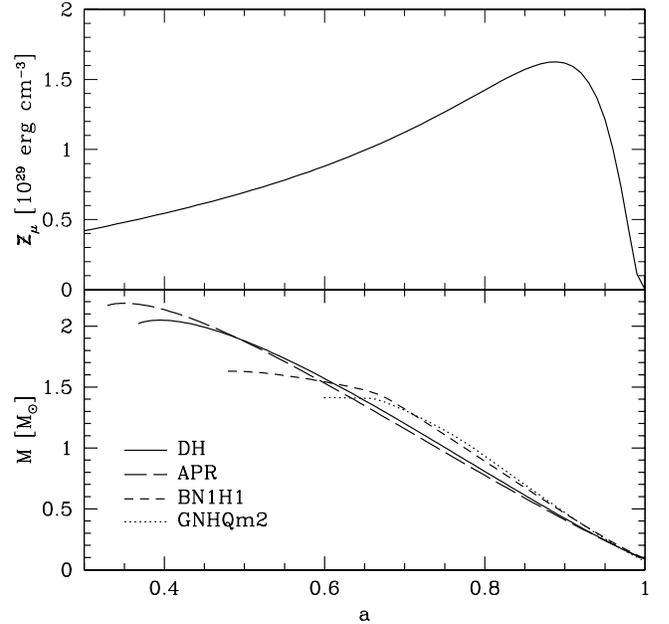}}
\caption{{\bf Upper panel:} the function ${\cal Z}_\mu(a)$, Eq.\
(\ref{eq:I_f}),
 calculated for the DH model of the ground-state crust, assuming
 $\mu(P)=
\mu(P)$, Eq.\ (\ref{eq:mu_eff}).
{\bf Lower panel:} relation between gravitational mass $M$ and
  $a=1-r_{\rm g}/R$, calculated for four EOSs of the neutron-star
  core.}
\label{fig:Int-a-M}
\end{figure}

\section{Calculation of $B$}
\label{sect:B}
We assume a quadrupolar deformation of the star.
Starting from a relaxed rotating configuration ${\breve {\cal C}}$,
characterized by some ${\breve \Omega}$ and ${\breve \epsilon}$,
 we calculate the hydro-elastic equilibrium
configuration ${\cal C^{\rm sd}}$ after a continuous, complete spin
down to $\Omega=0$.
The  configuration ${\cal C^{\rm sd}}$ is 
characterized by $\epsilon_{\rm r}>0$.
Within our
approximation, the hydro-elastic equilibrium at $\Omega=0$ implies
\begin{equation}
b={\epsilon_{\rm r}\over {\breve \epsilon}}~,~~B={b\over 1-b}A\approx bA.
\label{eq:B-epsilon_r}
\end{equation}

Classical analytic result, valid for a self-gravitating
 elastic sphere of constant $\rho$ and $\mu$, and volume $V$,
  is \citep{Love1920}
\begin{equation}
B={57\over 50} V \mu~. \label{eq:B-Love}
\end{equation}
In this case the displacement field $\pmb{u}(\pmb{r})$ is calculated
analytically \citep{Love1920}. \citet{BP1971} used this
displacement field to calculate $B=B_{\rm BP}$ for a compressible
solid crust (model of \citealt{BPS}),  enveloping a liquid core with
EOS of \citet{BBP}. \citet{PandhapPS1976} performed calculations of $B$
using two approximations. First, they used the prescription of
\citet{BP1971}. Second, they calculated $B$ using a modification
of Eq.\ (\ref{eq:B-Love}) obtained by replacing $57/50$ by one, $V$
by crust volume $V_{\rm c}$, and $\mu$ by the average over the
crust volume, $\overline{\mu}$. Therefore, their approximation
reads as
\begin{equation}
B\approx B_{\rm int}=\int_{\rm crust}{\rm d}^3\;r\mu(r)~.
\label{eq:B.int}
\end{equation}
\citet{PandhapPS1976}  found that $B_{\rm int}$ is a rather
good approximation of $B_{\rm BP}$.

As shown by \citet{Cutler2003}, the use of $\pmb{u}(\pmb{r})$ of
constant $\rho$ and $\mu$ solid sphere for a thin solid shell
enveloping a fluid core badly overestimates $B$. Using the analytic
$\pmb{u}(\pmb{r})$ of \citet{Franco2000} for an incompressible
shell, \citet{Cutler2003} derived, within the Cowling approximation,
a correct formula for a thin crust model of constant $\rho$ and $\mu$,
\begin{equation}
B_{\rm inc}={12\over 55} V_{\rm c} \mu~. \label{eq:Cutler-inc}
\end{equation}
The prefactor is smaller by 5.23 than in Eq.\
(\ref{eq:B-Love}). However, the use of Eq.\ (\ref{eq:Cutler-inc})
with a simple substitution $\mu\longrightarrow \overline{\mu}$ would
still badly overestimate the true value of $B$, by a factor $\approx
8$ for a $1.4~M_\odot$ star \citep{Cutler2003}. This is due to
additional cancellations of different contributions of elastic
strain to $\epsilon_{\rm r}$ \citep{Cutler2003}. The calculations
of \citet{Cutler2003} were done using the model of the crust of
\cite{NV1973}, combined with several EOSs of the core. Numerical
results, presented in Table 1 of \citet{Cutler2003}, were
obtained for a $1.4~M_\odot$ neutron star.
\begin{figure}[h]
\centering
\resizebox{3.5in}{!}{\includegraphics[clip]{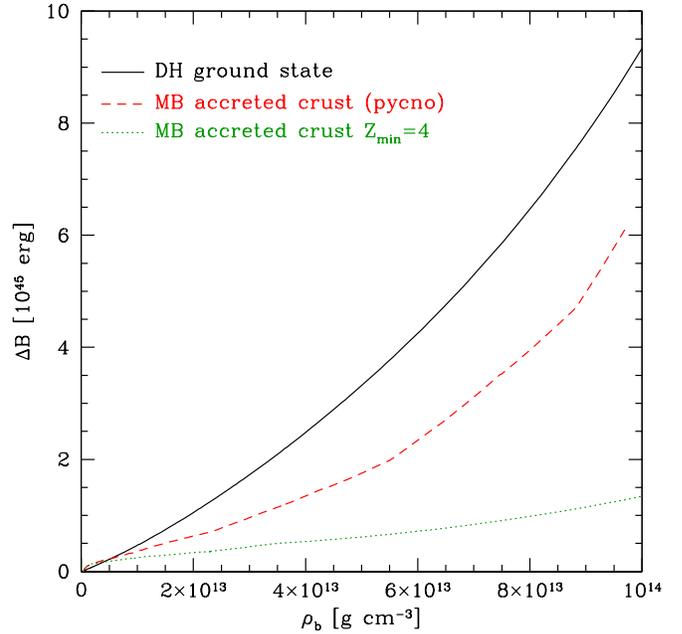}}
\caption{(Color online) Function  $\Delta B(\rho_{\rm b})$
calculated for three models of neutron-star crust
and assuming $a=0.65$. This value of $a$ is
obtained for a $1.4~M_\odot$ star model based
on the DH EOS for the crust and core (\citealt{DH2001}).
Two MB models differ in the treatment of the
pycnonuclear fusion. In both MB models, X-ray ashes are pure
$^{56}{\rm Fe}$. Green dotted line: artificial
blocking of pycnonuclear fusion
until $Z_{\rm min}=4$. Red dashed line: no artificial
suppression of pycnonuclear reactions, Table A.3 of Appendix
 to (\citealt{HZ2008}).  }
\label{fig:DeltaB}
\end{figure}

We do not attempt to find $B$ via solving the hydro-elastic
equilibrium equations, like in \citet{Cutler2003}. Instead, we
approximate $B$ by a generalization of Eq.\ (\ref{eq:B.int}) in the
form
\begin{equation}
B=0.027\int_{\rm crust}{\rm d}^3 r\;  \mu(r).
\label{eq:B-c_B}
\end{equation}
where the prefactor  accounts for the real distribution of strain
stresses within the crust after the spin down. Assuming  a reduction
factor $0.027\approx 1/42$ as compared to the Baym-Pines' estimate,
we reproduce results of \citet{Cutler2003} quite well.

\section{ $\Delta B(\rho_{\rm b})$ for the ground-state
 and accreted crust}
\label{sect:B-GS-ACC}

\begin{figure}[h]
\centering
\resizebox{3.5in}{!}{\includegraphics[clip]{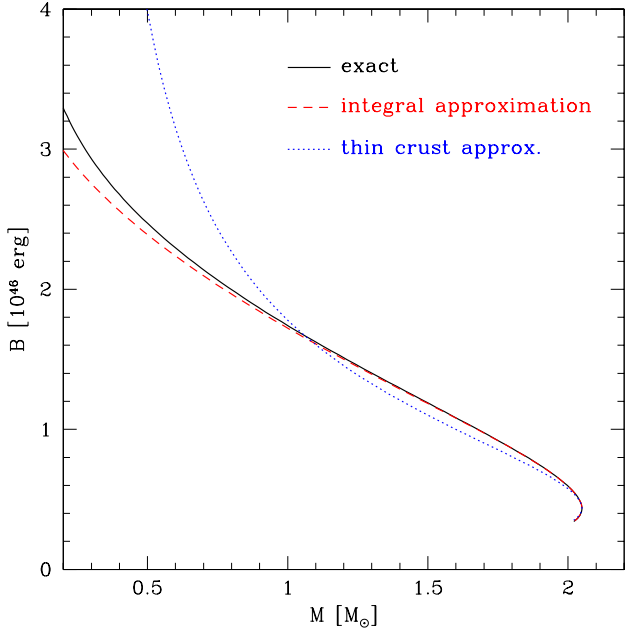}}
\caption{(Color online) Parameter $B$ versus stellar mass $M$
calculated for the DH EOS (Table 1). Solid line: exact result
obtained using Eq.\ (\ref{eq:B-c_B}). Dashed line: approximation
explained in Sect.\ \ref{sect:crust.approx}. Thin dotted line:
scaling with $M$ and $R$, as derived by \cite{Jones2000},
normalized here to the exact value of $B$ at $M=M_{\rm max}$.}
\label{fig:B_M}
\end{figure}
We denote by $\Delta B(\rho_{\rm b})$ the contribution to $B$ of the
crust layer with density $\rho_{\rm b}$ at the bottom. We use the approximation
given by Eqs.\ (\ref{eq:B-c_B}) and (\ref{intc}).
Therefore, the value of $\Delta B(\rho)$ is determined by
two functions, crust EOS and $\mu(P)$, and by
a single, star-structure parameter $a=1-r_{\rm g}/R$.
The dependence of $\Delta B(\rho)$
on the crust model is illustrated in  Fig.\
\ref{fig:DeltaB}. Three curves were obtained using different
crust models. The upper curve was based on the ground-state
model of \cite{DH2001}. The intermediate curve was calculated
assuming a model of accreted crust from \cite{HZ2008} (see
caption to Fig.\ \ref{fig:DeltaB}). The
values of $\mu$ are lower than for the ground-state crust, 
making $\Delta B(\rho_{\rm b})$ lower, too.

Pycnonuclear reaction rates at $\rho>10^{12}~\mdens$ are the
most uncertain element of an accreted crust model. An uncertainty
factor of $\sim 10^6$  is common for theoretical calculations
of reaction rates (\citealt{YakovlevGW-2006}). To visualize
the r{\^o}le of pycnonuclear fusion in determining $\Delta B$,
we (artificially) suppressed these reactions for $Z>Z_{\rm
min}=4$. Such modification has no significant effect on the
total deep crustal heating  in accreting neutron star
(\citealt{HZ2008}). However, as seen in Fig.\
\ref{fig:DeltaB}, it has a strong effect on $\Delta
B(\rho_{\rm b})$, decreasing it by a factor of two at $2\times
10^{13}~\mdens$ and by a factor of five at $10^{14}~\mdens$.
By nearly doubling $Z$ and decreasing
the number of nuclei by two in each reaction shell,
pycnonuclear reactions, 
have a very strong impact on the crustal rigidity.
\section{Scaling of $B$ with $M$ and $R$}
\label{sect:B_MR}
\cite{Jones2000} derived, within Newtonian gravity,
 a simple scaling of $\int_{\rm
crust}\mu{\rm d}V$ with $M$ and $R$.
The starting point was an estimate of the crust thickness
$\Delta R$, in the limit of $\Delta R/R \ll 1$.
By balancing gravity and pressure at the bottom of the crust,
one gets
\begin{equation}
\Delta R \propto R^2/M~.
\label{eq:DeltaR_Jones}
\end{equation}
In the same approximation, the crust volume scales
\begin{equation}
\Delta V=4\pi R^2\Delta R \propto R^4/M~.
\label{eq:Vcrust_Jones}
\end{equation}
Defining a mean value of the shear modulus, averaged over
the crust volume, $\left<\mu\right>_{\rm crust}$,  one gets
(\citealt{Jones2000})
\begin{equation}
B_{\rm tc}\propto \Delta V \left<\mu\right>_{\rm crust}
\propto R^4/M~,
\label{eq:B_tc}
\end{equation}
where the index ``tc'' recalls that the formula is obtained in
the limit of a {\bf t}hin {\bf c}rust. In what follows, we
 ``normalize'' the prefactor in the formula for $B_{\rm
tc}$ by  requiring that the formula reproduces  (for a fixed
EOS for the core {\it and} the crust) the value obtained by
an exact  integration $\int_{\rm crust}\mu{\rm d}V$ (see
the end of this section).

In Sect.\ \ref{sect:crust.approx} we derive an approximate,
but as we will show very precise,
 formula for $\int_{\rm crust}\mu{\rm d}V$.
In particular, we found that for a given $\mu(P)$ and EOS of
the crust, the relevant integral over the crust scales in a
well-defined manner with $R$ and $M$, the latter
scaling entering through the dependence on $a=1-r_{\rm g}/R$.
If we change $M,R\longrightarrow M^\prime,R^\prime$, we get,
in our approximation
\begin{equation}
B\longrightarrow B^\prime = (R^\prime/R)^3
{\cal Z}_\mu(a^\prime)/{\cal Z}_\mu(a)~,
\label{eq:scale.R.a}
\end{equation}
where  ${\cal Z}_\mu(a)$ is given by
 Eq.\ (\ref{eq:I_f}).

To be specific, let us consider the DH EOS of the crust and
the core, for which we also calculated $\mu=\mu(P)$. We
fix the prefactor in $B_{\rm tc}$ by requiring $B_{\rm tc}=B$
at $M=M_{\rm max}$, where the crust is the thinnest. In Fig.\
\ref{fig:B_M} we compare $B(M)$ obtained using three
approximations. The method based on Eq.\ (\ref{intc}) is very
precise, even for $M\approx 0.3~M_\odot$ when the crust is no
longer thin, while for $M>0.5~M_\odot$ its precision is
remarkably high. Approximation $B_{\rm tc}$ is reasonably good
only for $M>M_\odot$ and
provided the value of $\left<\mu\right>_{\rm crust}$ is properly chosen,
which requires a ``normalization''
by an exact value of the integral at a selected value of $M$.
For $M<M_\odot$, $B_{\rm tc}$ diverges badly from the precise
$B(M)$, and the scaling given by Eq.\ (\ref{eq:B_tc}) cannot be
used.
\section{Numerical results for crustal rigidity $b$}
\label{sect:num-res-b}
\begin{figure}[h]
\centering
\resizebox{3.5in}{!}{\includegraphics[clip]{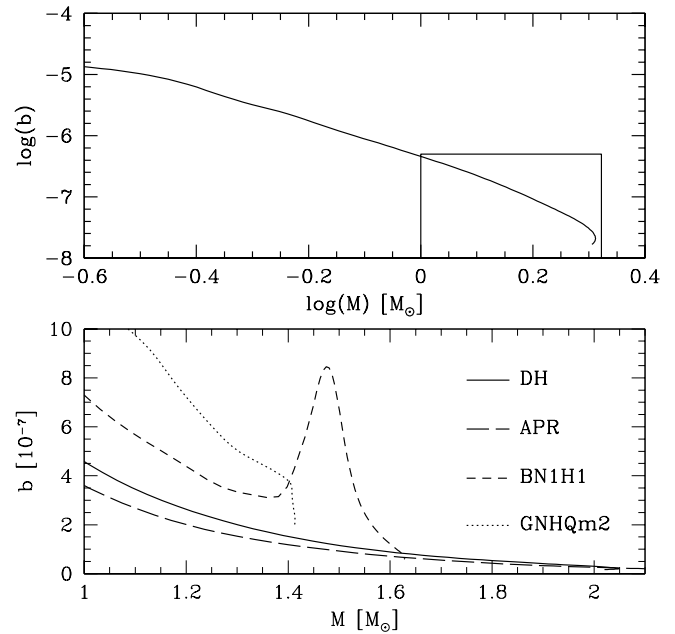}}
\caption{Crustal rigidity $b$ versus stellar mass.
{\bf Upper panel:} log-log
plot, which shows results for the DH model
of the ground-state crust and core  for $0.2~M_\odot<M<M_{\rm max}$.
{\bf Lower panel:} zoomed segment of $b(M)$ for $M>1~M_\odot$
(within a rectangle in the lower-right corner of the upper panel)
for the DH model
of the ground-state crust combined with several EOSs of the core.}
\label{fig:b-M-log-lin}
\end{figure}

Our results are presented  in Fig.\ \ref{fig:b-M-log-lin}. In
the upper panel  we show our results for $b(M)$ calculated
using the DH model of the crust and core, for $0.25~M_\odot
<M<M_{\rm max}({\rm DH})=2.05~M_\odot$. In the lower panel we
have plotted $b(M)$ for $M>1~M_\odot$, calculated using the DH
model of the crust and four different EOSs of the core.
\subsection{Nucleon cores}
\label{sect:num-Ncores}
As examples, we used the DH and APR EOSs for the core. The core
EOS does not exhibit  any strong softening due to
hyperonization or a phase transition, so $b(M)$ is
monotonous. It increases with decreasing $M$ because of
a monotonous decrease in $A$ (Fig.\ \ref{fig:A_EOS1})  and
an increase in $B$ due to an monotonous increase in the crust volume.
Crustal rigidity is a sensitive function of $M$; the value of
$b$  at $M=1.0~M_\odot$ is ten times greater than at
$M=2~M_\odot$. For a ``standard'' mass $1.4~M_\odot$ we get
$b\approx 2\times10^{-7}$, which is within the range of $b=1.5-4.4\times
10^{-7}$, obtained  by \cite{Cutler2003} for the FPS crust
(\citealt{Lorenz1993}) and
Negele-Vautherin crust (\citealt{ NV1973}), each combined with two different
EOSs of the core (see Table 1 of \citealt{Cutler2003}).

The increase in $b$ with decreasing $M$ is so strong that
$b(0.3~M_\odot)\approx 50 b(1.4~M_\odot)$ and we were forced
to use the logarithmic scale to plot $b$ in the whole
considered mass range. Even at
$M=0.3~M_\odot$, the maximum elastic strain effect on neutron
star structure, measured by $b\sim 10^{-5}$, is a very small
perturbation of the perfect-fluid model. However, one also has
to notice that the applicability of the small deformation
approximation to neutron stars with $M<1~M_\odot$, rotating
at a few  hundred Hz, can be questioned.

Consider, for example, $f=641~$Hz, corresponding to PSR
B1937+26. For illustration, we use properties of the
low-mass neutron stars calculated for the DH EOS of the crust
and core  by \cite{HZD2002}. At $f=641~$Hz, the equatorial
radius of a $0.6~M_\odot$ star is $17.8$ km, to be compared
with $12.6~$km for a nonrotating star of the same mass. The
polar flattening is characterized  by $r_{\rm pole}/r_{\rm
eq}\approx 0.6$ (here $r$ is the radial coordinate). The
equatorial thickness of the crust is four times that at the
poles. Actually, at $f=641~$Hz, stars with $M<0.6~M_\odot$
cannot exist in hydrostatic equilibrium:
 they are destroyed due to the mass shedding from the equator.

 If DH EOS of the core is replaced by the APR one, the change
 in $b(M)$ is small. Namely, $b(M)$ is  somewhat lower, due to
 a higher stiffness of the APR EOS.
\subsection{Hyperon cores}
\label{sect:num-Hcores}
A strong softening of the core EOS due to hyperonization has a
dramatic effect on the high-$M$ segment of $b(M)$. Namely, the
crustal rigidity is no longer monotonous, and after reaching a
rather narrow maximum of nearly $10^{-6}$ at $\sim
1.5~M_\odot$, it falls monotonously, down to about $10^{-7}$,
at $M=M_{\rm max}({\rm BGN1H1})=1.65~M_\odot$. This behavior
is easily understood in terms of a specific dependence of $A$
on $M$ (see Fig.\ \ref{fig:A_EOS2}), characterized by a deep
and narrow  minimum around  $\sim 1.5~M_\odot$. A minimum  in
$A(M)$ is reflected by a maximum in $b(M)$.

\subsection{Exotic cores}
\label{sect:num-Exot-cores}
Exotic cores are represented by the GNHQm2 EOS, with a phase
transition to quark matter via a mixed quark-hadron state. The
softening effect implies  a low value of $M_{\rm max}$, high
compactness of ``hybrid stars''  and rapid shrinking of the
crusts with $M$ approaching $M_{\rm max}$. This results in a
rapid drop in $b$ near $M_{\rm max}$. The effect of a
softening due to mixed state is seen  in Fig.\
\ref{fig:A_EOS2}. It is represented by flattening of the
$A(M)$ curve GNHQm2, and it is more gradual than a
violent softening by a hyperonization, as seen in curve BGN1H1 in
the same figure.

\section{Discussion and conclusion}
\label{sect:discussion}
In the present paper we have studied the dependence of the $A$ and
$B$ parameters of the Baym-Pines model of the hydro-elastic
equilibrium of rotating neutron stars on the stellar mass and
on the EOS. We considered seven different EOSs of the liquid
core and several  models of neutron star crust. \vskip 2mm
\parindent 0pt

{\it Parameter $A$.} It was calculated using high-precision
2-D simulations in General Relativity, so it can be
considered as exact, in contrast to the previous estimates
based on the incompressible fluid model. We developed a
systematic procedure for calculating $A(M)$. This procedure
 is based on linear expansions of global parameters of rotating
 neutron stars of a fixed baryon mass in the square of
 rotation frequency as measured by an observer at infinity.
 Three different coefficients of the quadratic term
  measure, respectively, the anisotropy of the inertia tensor
  ($\alpha_I$),   oblateness of shape ($\alpha_{\rm s}$),
  and increase in the energy as measured in the star's rest
  frame ($\alpha_{\rm d}$).  Combined, they give
  $A=\alpha_{\rm d}(\alpha_{\rm  s}/\alpha_I)^2$. A precise
  calculation of $A(M)$ was possible for $0.2\;\msun\la M \la
  0.9\;M_{\rm max}$. Our exact values of $A(M)$ are
  usually  much   higher than those given by the
  standard formula based on the incompressible fluid model,
  used in the previous calculations.
  For $M<0.7\;\msun$ the values of $A(M)$ are determined by
  a low-density segment of the EOS (similar for all EOSs) and
  are given by a universal formula of precision on the order
  of one percent.  For $M>0.7\;\msun$, values of $A(M)$
  are EOS dependent and diverge
  with increasing $M$. For EOSs without a high-density softening
  (only nucleons and leptons),  the values $A(M)$ increase
  monotonously with $M$. For EOSs with a high-density
  softening (due to hyperons or a phase transition), $A(M)$ can
  have a strongly non-monotonous high-mass segment, where
  $A$ can drop strongly  with an increasing $M$. For example,
  softening by hyperons in the BGN1H1 EOS makes $A$
  drop by a factor of about three for $M$ increasing from $1.4\;\msun$
  to $1.5\;\msun$. Generally,  the EOS dependence of
  $A(M)$ for $M\approx 1.5\;\msun$ is
  very strong, and  $A$ can increase by a factor of five  when going
   from  hyperon-softened EOS to a stiff nucleonic EOS of the
   APR type.

\vskip 2mm

 {\it Crustal rigidity.} We derived an
approximate formula for the integral of the elastic strain
over the crust volume, whose validity  goes beyond  a thin
crust approximation. Our formula can also be used for low-mass
neutron stars. We considered three different models of the
crust, two for the ground-state crust and one for an accreted
crust. The values of $B$ for different models of the
ground-state crust and the same core EOS show little model
dependence. In contrast, the accreted crust is
significantly less rigid (lower $B$) than the ground-state
one. Results obtained for $M=1.4~M_\odot~$ with the
ground-state  DH crust and nucleonic cores are within the
range of values obtained by \cite{Cutler2003}; however, we
found a strong dependence of $b$ on the star's mass. In the
range $1.0~M_\odot - M_{\rm max}$, parameter $b$ varies by an
order of magnitude, and one notices a characteristic
dependence of relation $b(M)$ on the EOS of the core. For
nucleonic cores, $b(M)$ is monotonic. However, hyperonization
may lead to a strong maximum in $b(M)$, associated with soft
hyperon cores and a deep minimum in $A(M)$. A phase transition
via a mixed quark-hadron state implies a strong drop in $b$.

\vskip 2mm

{\it Pycnonuclear reactions and crustal rigidity.} We have shown
that pycnonuclear reactions, if they proceed,
 play a crucial r{\^o}le for rigidity of
accreted crust, because every time they occur they nearly double
$Z$. However, theoretical pycnonuclear reaction rates are notorious
for their huge uncertainties. There is also a basic difficulty
in describing of aggregates of neutrons, bound to the proton
clusters, bound in turn in the nuclear potential well created by
neutrons, as ``nuclei'' immersed in a superfluid of dripped
(unbound) neutrons. The effect  of dripped neutrons on the
zero-point vibrations of such ``nuclei'' and on the fusion rate
still needs to be calculated. We use a compressible liquid drop model
of nuclei in the inner crust. 
 Assuming that pycnonuclear fusions
are suppressed for $Z>Z_{\rm min}=4$,
we get $B$, which is five times smaller than for a
model without such an artificial suppression. This may be
contrasted with deep crustal heating in accreting neutron
stars, where the total heat release per one accreted nucleon
was found to be insensitive to the suppression of pycnonuclear
reactions (\citealt{HZ2008}).

\vskip 2mm
\parindent 21pt

This paper is first in a series devoted to the interplay of
rotation and shear strain in neutron stars. The forthcoming
papers will deal with hydro-elastic equilibrium of rotating
neutron stars, including the case of hypothetical solid
neutron-star cores, predicted by some theories of dense
matter. Elastic strain can support non-axial mass
distributions of rotating neutron stars, therefore becoming
sources of continuous, strictly periodic gravitational waves.
Such models of gravitational wave sources are of particular
interest in the expected   era of gravitational-wave
astronomy, which could be (hopefully) opened  by LIGO, VIRGO,
and other gravitational-wave observatories in the coming
decade.

\acknowledgements{ This work was partially supported by the
Polish MNiSW grant no.
N20300632/0450, and by the LEA Astrophysics Poland-France (Astro-PF) program.
MB was also supported by the Marie Curie Fellowships no. MEIF-CT-2005-023644
and ERG-2007-224793}


\end{document}